\documentclass{article}

\usepackage{arxiv}

\usepackage[utf8]{inputenc} 
\usepackage[T1]{fontenc}    
\usepackage{hyperref}       
\usepackage{url}            
\usepackage{booktabs}       
\usepackage{amsfonts}       
\usepackage{nicefrac}       
\usepackage{microtype}      
\usepackage{lipsum}
\usepackage{graphicx}
\usepackage{epstopdf}
\usepackage{amsmath}
\usepackage{psfrag}
\usepackage{epsfig}
\title{Improved Multi-step FCS-MPCC with Disturbance Compensation for PMSM Drives -- Methods and Experimental Validation
}

\author{
Hai Yang \\
  College of Information Engineering\\
  Zhejiang University of Technology\\
  Hangzhou, China.\\
  \texttt{yanghai19@163.com} \\
   \And
 Yibin Liu \\
  College of Information Engineering\\
  Zhejiang University of Technology\\
  Hangzhou, China.\\
  \texttt{liuyibin2020@163.com} \\
  \And
 Junxiao Wang \\
  College of Information Engineering\\
  Zhejiang University of Technology\\
  Hangzhou, China.\\
  \texttt{wjx2017@zjut.edu.cn} \\
  \And
Jun Yang \\
  Department of Aeronautical and Automotive Engineering\\
  Loughborough University\\
  Loughborough LE11 3TU, U.K.\\
  \texttt{j.yang3@lboro.ac.uk} \\
}

\begin{document}
\maketitle
\begin{abstract}
In this paper, an improved multi-step finite control set model predictive current control (FCS-MPCC) strategy with speed loop disturbance compensation is proposed for permanent magnet synchronous machine (PMSM) drives system. A multi-step prediction mechanism is beneficial to significantly improve the steady-state performance of the motor system. While the conventional multi-step prediction has the defect of heavy computational burden, an improved multi-step finite control set model predictive current control (IM MPCC) strategy is proposed by developing a new multi-step prediction mechanism. Furthermore, in order to improve the dynamic response of the system, a disturbance compensation (DC) mechanism based on an extended state observer (ESO) is proposed to estimate and compensate the total disturbance in the speed loop of the PMSM system. Both simulation and experimental results validate the effectiveness of the proposed control strategy.
\end{abstract}


\section{Introduction}
Permanent magnet synchronous motor (PMSM) has the advantages of small size, high power density, high ratio of moment of inertia and high operating efficiency, which admit it widely used in electric vehicles, ships, railway transportation and industrial robot and other fields \cite{paper1}-\cite{paper2}. However, PMSM drive system is essentially a complex nonlinear multivariable coupling system, which is quite sensitive to system parameter uncertainty and external disturbance, and thus make it difficult to meet the increasing control requirements with respect to higher-precision serving \cite{paper3}. Traditional linear control strategies, having a slow response time under parameter changes and external disturbances, would lead to unsatisfactory dynamic performance of the system \cite{paper4}-\cite{paper6}. In order to solve these problems, many new nonlinear control technologies have been investigated and applied to the PMSM drives system, such as sliding model control \cite{paper7}, model reference adaptive control \cite{paper8}, active disturbance rejection control \cite{paper9}, model predictive control (MPC) \cite{paper10}, etc.

Because of simple implementation, straightforward handling of nonlinearities and constraints, and good dynamic
response, the MPC has attracted more attention in recent years. MPC can be divided into continuous control set model predictive control (CCS-MPC) \cite{paper11}-\cite{paper12} and finite control set model predictive control (FCS-MPC) \cite{paper13}. Compared with CCS-MPC, FCS-MPC dose not require complex pulse width modulation, which directly manipulates the inverter with switching signals, delivering much faster dynamic performance of power converters and motor drives  \cite{paper14}.


Since FCS-MPC is based upon a certain number of control sets, there must be a defect that large ripples lead to unsatisfactory steady-state performance \cite{paper15}. In order to reduce the torque and magnetic flux ripple, in \cite{paper16}, two non-zero vectors are applied in one control cycle to obtain better steady-state performance. In \cite{paper17}, a passivity-based model predictive control strategy is proposed, which improves the steady-state performance of the system while reducing the computational burden.
T. Geyer et al. \cite{paper18} propose a modified sphere decoding algorithm, which can solve the optimization problem underlying MPC for a longer prediction horizon. However, when it comes to the case of multi-level converters, the spherical decoding algorithm still shows its limitation in computational efficiency. Therefore, an improved spherical decoding algorithm is proposed in \cite{paper19}, which makes direct MPC with long prediction horizon an attractive and computationally affordable control scheme. It should be highlighted that the complexity of this algorithm still brings a barrier for practical implementation. These methods are improved the computational efficiency as compared with the baseline multi-step FCS-MPC, but the control system is still vulnerable to the system parameters uncertainty and external disturbances \cite{paper20}. 

In order to suppress the system parameters uncertainty and external disturbances for better dynamic response of the system, the disturbance estimation technology is gradually introduced into the field of high performance control of PMSM drives \cite{paper21}-\cite{paper23}. In \cite{paper1}, the predictive function control method (PFC) is combined with the extended state observer (ESO) to obtain the optimal control rate by minimizing the quadratic performance index, which has certain anti-disturbance effect. But in the case of strong disturbance, the result is not satisfactory.


Motivated by the above-mentioned challenges, an improved multi-step FCS-MPCC strategy with speed loop disturbance compensation is proposed to achieve improved steady-state performance and dynamic response for PMSM system. The design philosophy of the proposed strategy is explicitly demonstrated in the following two phases. First, based on a single-step prediction, the optimal voltage vector and the sub-optimal voltage vector calculated from the cost function are utilized as the vector waiting to be applied. The vector applied to the inverter is determined according to the range of the optimal voltage vector which predicted in the second step. Second, in order to improve the dynamic response of the system, a disturbance compensation mechanism based on an ESO is added to estimate and compensate the total disturbance for the speed loop.

The structure of this paper is as follows: Section 2 introduces the mathematical model of PMSM and the conventional multi-step model. Section 2 explains the proposed control strategy in this paper. The simulation and experimental results are shown in Section 4 and Section 5 respectively. The conclusion of proposed strategy are drawn in Section 6.

\section{ Conventional  FCS-MPCC}

\subsection{PMSM  Model}
The mathematical model of PMSM can be described as follows:
\begin{align}
\label{1} u_d&= {R_s}{i_d} + {L_d}\frac{{d{i_d}}}{{dt}} - {\omega _{re}}{L_q}{i_q}\\
\label{2} u_q&= {\omega _{re}}{L_d}{i_d}{\rm{ + }}{R_s}{i_q} + {L_q}\frac{{d{i_q}}}{{dt}} + {\omega _{re}}{\psi _f}\\
\label{3} J\frac{{d\omega }}{{dt}}{\rm{ }}&= \frac{3}{2}{p_n}({\psi _f}{i_q} + ({L_d} - {L_q}){i_d}{i_q}) - B\omega  - {T_L}
\end{align}

where$\ u_{d}$,$\ i_{d}$,$\ L_{d}$ represent d axis voltage, current, inductance respectively. $\ u_{q}$,$\ i_{q}$,$\ L_{q}$ represent q axis voltage, current, inductance respectively.$\ R_{s}$,$\ p_{n}$,$\ \psi_{f}$ represent the stator resistance, the number of pole pairs, rotor flux respectively. $\ \omega_{re}$, $\ \omega$, $\ J$, $\ B$ and$\ T_{L}$ represent electrical angular velocity, mechanical angular velocity, moment of inertia, friction coefficient, load torque respectively.

\subsection{Single-step FCS-MPCC}
Compared with the system which adopts FOC strategy, the cost function which is based on the error is introduced to the current internal loop instead of PI controller. The numbers of the voltage vectors in the finite control set depend on the selected inverter, and the optimal voltage vector is obtained by minimize the cost function, which can be applied to the inverter directly.

According to the model (1),(2) the differential form of the current is obtained by discretizing the current state equation by the forward Euler method.
\begin{align}
&\begin{array}{l}
\label{4}{i_d}(k + 1) = \left( {1 - {R_s}G} \right){i_d}(k) + {L_q}G{\omega _{re}}(k){i_q}(k)\\
\;\;\;\;\;\;\;\;\;\;\;\;\;\;\;\;\; + G{u_d}(k)
\end{array}\\
&\begin{array}{l}
\label{5}{i_q}(k + 1) =  - {L_d}H{\omega _{re}}(k){i_d}(k){\rm{ + }}\left( {1 - {R_s}H} \right){i_q}(k)\\
\;\;\;\;\;\;\;\;\;\;\;\;\;\;\;\;\; + H{u_q}(k) - {\psi _f}H{\omega _{re}}(k)
\end{array}
\end{align}

where $\ T_{s}$ is the sample time, $\ G = \frac{{{T_s}}}{{{L_d}}}$,$\ H = \frac{{{T_s}}}{{{L_q}}}$.

 Because there exists the computation delay in the actual system, the optimal voltage vector at the present sampling period will be applied to the inverter at the next sampling period,which will influence the system performance.It is necessary to make one-step delay compensation for the system. Equation (4),(5) can be used to compensate one step before the model prediction,then the predicted current value and cost function value can be obtained by putting diferent voltage vector from the finite voltage vector control set $\ ({u_{fv}})$ into the following formular:
 \begin{align}\label{6}
&\begin{array}{l}
i_d^{pre}(k + 1) = \left( {1 - {R_s}G} \right){i_d}(k + 1) + Gu_{fv}^d\\
\begin{array}{*{20}{c}}
{}&{}&{\begin{array}{*{20}{c}}
{}&{}
\end{array}}
\end{array} + {L_q}G{\omega _{re}}(k + 1){i_q}(k + 1)
\end{array}\\
\label{7}&\begin{array}{l}
i_q^{pre}(k + 1) =  - {L_d}H{\omega _{re}}(k + 1){i_d}(k + 1) + Hu_{fv}^q\\
\begin{array}{*{20}{c}}
{\begin{array}{*{20}{c}}
{}&{}
\end{array}}&{}&{}
\end{array}{\rm{ + }}\left( {1 - {R_s}H} \right){i_q}(k + 1) - {\psi _f}H{\omega _{re}}(k + 1)
\end{array}\\
\label{8}&\begin{array}{l}
g = \left| {i_d^* - i_d^{pre}(k + 1)} \right| + \left| {i_q^* - i_q^{pre}(k + 1)} \right|\\
\;\;\;\;\;\;\;\;\;\;\;\;+ \lim (i_{d,q}^{pre}(k + 1))
\end{array}
\end{align}

where, ${i^*}$ is current reference. In formula (8), the first term and second term evaluate the predicted current error, the last term is the current constraint condition, which can be expressed as
 \begin{align}
\label{9}\lim (i_{d,q}^{pre}(k + 1)) = \left\{ \begin{array}{l}
0,\;if\;i_d^{pre}(k + 1) \le {i_{\max }}\;
\\\;\;\;\;and\;i_q^{pre}(k + 1) \le {i_{\max }}\\
\infty ,if\;i_d^{pre}(k + 1) > {i_{\max }}\;
\\\;\;\;\;or\;i_q^{pre}(k + 1) > {i_{\max }}
\end{array} \right.
\end{align}

where $i_{\max }^{}$ is the maximum instantaneous current.
\subsection{Conventional Multi-step FCS-MPCC}
The prediction time domain can be extended to N step on the basis of single-step FCS-MPCC, the following equations (10),(11),(12) can be obtained:
\begin{align}
\label{10} &\begin{array}{*{20}{c}}
{i_d^{pre}(k + N) = \left( {1 - {R_s}G} \right){i_d}(k + N) + Gu_{fv}^d}\\
{\begin{array}{*{20}{c}}
{\begin{array}{*{20}{c}}
{}&{}
\end{array}}&{}&{}
\end{array} + {L_q}G{\omega _{re}}(k + N){i_q}(k + N)}
\end{array}\\
\label{11} &\begin{array}{l}
i_q^{pre}(k + N) =  - {L_d}H{\omega _{re}}(k + N){i_d}(k + N) + Hu_{fv}^q\\
\begin{array}{*{20}{c}}
{\begin{array}{*{20}{c}}
{}&{}&{}&{}
\end{array}}&{}
\end{array}{\rm{ + }}\left( {1 - {R_s}H} \right){i_q}(k + N) - {\psi _f}H{\omega _{re}}(k + N)
\end{array}\\
\label{12}&g = \sum\limits_{n = 1}^N {\left[ {\begin{array}{*{20}{l}}
{\left| {i_d^* - i_d^{pre}(k + n)} \right| + \left| {i_q^* - i_q^{pre}(k + n)} \right|}\\
{ + \lim (i_{d,q}^{pre}(k + n)}
\end{array}} \right]}\\
\label{13}&\lim (i_{d,q}^{pre}(k + n)) = \left\{ \begin{array}{l}
0,\;if\;i_d^{pre}(k + n) \le {i_{\max }}\;\\
\;\;\;and\;i_q^{pre}(k + n) \le {i_{\max }}\\
\infty ,if\;i_d^{pre}(k + n) > {i_{\max }}\;\\
\;\;\;\;or\;i_q^{pre}(k + n) > {i_{\max }}
\end{array} \right.
\end{align}

where$\ {n = 1,2, \cdots ,N}$,${N}$ is the predicted step size, the current limit condition is expressed as (13).
\begin{figure}[htbp]
\begin{center}
\psfrag{u3}{t3}
\psfrag{010}{8}
\includegraphics[height=3.5cm,width=4cm]{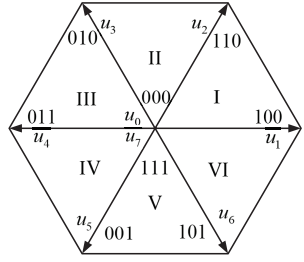}
\end{center}
\caption{The finite voltage vector control set.}
\label{1}
\end{figure}

Multi-step prediction needs to take the number of calculations into account, for example, when$\ {N=2}$, the controller has to calculate at least 72 times; with the increase of step size, the amount of calculation will increase exponentially, and when$\ {N=3}$, the controller needs at least 584 operations. Although the long prediction step size may bring better steady-state performance to the system, the algorithm is difficult to implement because of the limited processing capacity of the processor in the actual system.
\section{Improved Multi-step FCS-MPCC with speed disturbance compensation}
The voltage vector selected by the traditional FCS-MPC method is only optimal for one control cycle, and it is not optimal for multiple control cycles, which makes the steady-state performance of the system unsatisfactory.
As shown in Fig. 2, the current value after one step delay compensation at time $\ (k + 1)T$, the current prediction value of the finite control set is obtained through the prediction model at time $\ (k + 1)T_{pre}$, and the second step prediction is performed at time $\ (k + 2)T_{pre}$. It is easy to find that the control quantity closest to the reference value obtained at time $\ (k + 1)T_{pre}$ is not optimal at time $\ (k + 2)T_{pre}$. This problem illustrates that single-step prediction may make the system have unsatisfactory steady-state performance. The proposed strategy of IM MPCC in the paper can improve these shortcomings and has less computational burden.
\begin{figure}[htbp]
\begin{center}
\includegraphics[height=4cm,width=8cm]{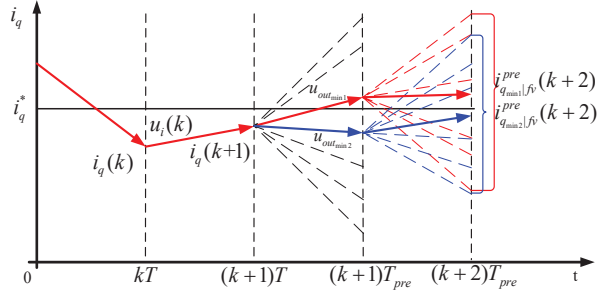}
\end{center}
\caption{Multi-step FCS-MPCC.}
\label{1}
\end{figure}

Furthermore, a disturbance compensation(DC) mechanism based on extended state observer (ESO) has been added to estimate and compensate the total disturbance of the speed loop to improve the dynamic response of the system. The proposed strategy is shown in Fig.3.
\subsection{Improved Two-step FCS-MPCC}\label{AA}
On the basis of single-step prediction, the eight voltage vectors are calculated by the prediction model equation (6, 7) and the cost function equation (8) to obtain the voltage vectors that minimize and subminimize the cost function. We call it the optimal output voltage vector $\ u_{out_{min1}}$ and the sub-optimal output voltage vector $\ u_{out_{min2}}$ at time $\ (k + 1)T_{pre}$.
\begin{align}\label{Mechanica5}
u_{ou{t_{\min 1}}} = \left[ {\begin{array}{*{20}{c}}
{{S_{a\min 1}}}\\
{{S_{b\min 1}}}\\
{{S_{c\min 1}}}
\end{array}} \right],\left\{ {\min 1({g_i}),i = 0,1...7} \right\}\\
\label{Mechanica6}
u_{ou{t_{\min 2}}} = \left[ {\begin{array}{*{20}{c}}
{{S_{a\min 2}}}\\
{{S_{b\min 2}}}\\
{{S_{c\min 2}}}
\end{array}} \right],\left\{ {\min 2({g_i}),i = 0,1...7} \right\}
\end{align}

Calculate the optimal and sub-optimal current prediction values at time  $\ (k + 2)T_{pre}$ through the prediction model, and express the prediction model as follow
\begin{align}\label{14}
&\begin{array}{l}
i_{{d_{\min 1}}}^{pre}(k + 2) = \left( {1 - {R_s}G} \right)i_{{d_{\min 1}}}^{pre}(k + 1) + Gu_{fv}^d\\
\begin{array}{*{20}{c}}
{\begin{array}{*{20}{c}}
{\begin{array}{*{20}{c}}
{}&{}
\end{array}}&{}
\end{array}}&{}
\end{array} + {L_q}G{\omega _{re}}(k + 2)i_{{q_{\min 1}}}^{pre}(k + 1)
\end{array}\\
\label{15} &\begin{array}{l}
i_{{q_{\min 2}}}^{pre}(k + 2) =  - {L_d}H{\omega _{re}}(k + 2)i_{{d_{\min 2}}}^{pre}(k + 1) + Hu_{fv}^q\\
\begin{array}{*{20}{c}}
{\begin{array}{*{20}{c}}
{}&{}
\end{array}}&{}&{}
\end{array}{\rm{ + }}\left( {1 - {R_s}H} \right)i_{{q_{\min 2}}}^{pre}(k + 1) - {\psi _f}H{\omega _{re}}(k + 2)
\end{array}
\end{align}

Compared with the conventional two-step prediction, the proposed improved two-step prediction will obtain 16 cost function values, reducing at least 48 model and cost function calculations. The new cost function select act voltage vector applied to inverter at time $\ (k+1)T_{pre}$ based on the range of optimal voltage vector predicted at time $\ (k+2)T_{pre}$.
\begin{align}
\label{Mechanica9}
{g_{sum}} = \left[ {\begin{array}{*{20}{c}}
A\\
B
\end{array}} \right] = \left[ {\begin{array}{*{20}{c}}
{g_{\min 1}^0}&{g_{\min 1}^1}& \ldots &{g_{\min 1}^7}\\
{g_{\min 2}^0}&{g_{\min 2}^1}& \ldots &{g_{\min 2}^7}
\end{array}} \right]
\end{align}
\begin{align}
\label{Mechanica9}
{V_{out}}{\rm{ = }}\left\{ {\begin{array}{*{20}{c}}
{{u_{ou{t_{\min 1}}}},if\begin{array}{*{20}{c}}
{\min ({g_{sum}}) \in A}
\end{array}}\\
{{u_{ou{t_{\min 2}}}},if\begin{array}{*{20}{c}}
{\min ({g_{sum}}) \in B}
\end{array}}
\end{array}} \right.
\end{align}

where, A and B respectively represent the set of cost function values at time $\ (k + 2)T_{pre}$ calculated based on the optimal voltage vector and the sub-optimal voltage vector, ${g_{sum}}$ is the set of all predicted cost function values at time $\ (k + 2)T_{pre}$, and ${V_{out}}$ is the voltage applied to the inverter in the end. If the minimum cost function value in ${g_{sum}}$ belongs to A set, the optimal voltage vector at time $\ (k + 1)T_{pre}$ is applied; if the minimum cost function value in ${g_{sum}}$ belongs to B set, the suboptimal voltage vector at time $\ (k + 1)T_{pre}$ is applied.
\begin{figure}[htbp]
\begin{center}
\includegraphics[height=4cm,width=8.5cm]{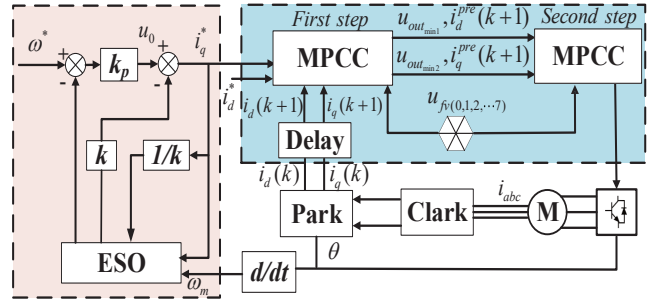}
\end{center}
\caption{IM MPCC strategy with speed loop DC.}
\label{1}
\end{figure}
This strategy can be topologically applied to the long prediction step size model. Compared with the conventional FCS-MPCC cost optimization method, this strategy has obvious advantages in less computation but similar steady-state performance. When the prediction steps  ${N=3}$, the number of calculations of the model and cost function is only 56 times. The final selection function in multi-step prediction can be described as
\begin{align}
\label{Mechanica11}
g_{sum} = {\left[ {\underbrace {\begin{array}{*{20}{c}}
A&B&C& \ldots
\end{array}}_{{2^{N - 1}}}} \right]^{\rm{T}}}
\end{align}
\begin{align}
\label{Mechanica12}
&V_{out} = \left\{ {\begin{array}{*{20}{c}}
{{u_{ou{t_{\min 1}}}},if\begin{array}{*{20}{c}}
{\min ({g_{sum}}) \in {{\left[ {\overbrace {\begin{array}{*{20}{c}}
A&B& \ldots
\end{array}}^{[1,\frac{{{2^{N - 1}}}}{2}]}} \right]}^T}}
\end{array}}\\
{{u_{ou{t_{\min 2}}}},if\begin{array}{*{20}{c}}
{\min ({g_{sum}}) \in {{\left[ {\overbrace {\begin{array}{*{20}{c}}
E&F& \ldots
\end{array}}^{(\frac{{{2^{N - 1}}}}{2},{2^{N - 1}}]}} \right]}^T}}
\end{array}}
\end{array}} \right.
\end{align}

 If the minimum cost function value in ${g_{sum}}$ belongs to set before the $\ \frac{{{2^{N - 1}}}}{2}$th, the optimal voltage vector at time $\ (k + 1)T_{pre}$ is applied; if the minimum cost function value in ${g_{sum}}$ belongs to other set in $\ \frac{{{2^{N - 1}}}}{2}$th to $\ 2^{N - 1}$th, the suboptimal voltage vector at time $\ (k + 1)T_{pre}$ is applied. Since the final voltage vector acting on the inverter is always the voltage vector at time $\ (k + 1)T_{pre}$, the effect is not as obvious when the prediction step is too long. so, the final step size is two.

\subsection{Speed Loop Disturbance Compensation}
Considering the uncertainty of system parameter uncertainty and external disturbance, the state variable
${d_\omega } =  - \frac{{B{\omega}}}{J} - \frac{{{T_L}}}{J} + \frac{3}{2}p{\psi _f}(\frac{{{i_q}}}{J} - \frac{{{i_q}^*}}{{{J_n}}})$ is introduced, ${J_{n} }$  is the moment of inertia of the rotor.If the sampling time is small enough, ${d_\omega }$ can be regarded as approximately unchanged, so let $\ {{\dot d}_\omega }   = 0$.

From this, the new equation of state can be deduced as
\begin{align}
\label{Mechanica19}
\left\{ {\begin{array}{*{20}{l}}
{{{\dot \omega }} = \frac{{{i_q}^*}}{k} + {d_\omega }}\\
{{{\dot d}_\omega } = 0}
\end{array}} \right.
\end{align}

According to the formula (22), the extended state observer (ESO) is designed as
\begin{align}
\left\{ {\begin{array}{*{20}{l}}
{{{\dot z}_1} = \frac{{{i_q}^*}}{k} + {z_2} - {\beta _1}({z_1} - {\omega})}\\
{{{\dot z}_2} =  - {\beta _2}({z_1} - {\omega})}
\end{array}} \right.
\end{align}

As long as the parameters $\ \beta_{1} $ and $\ \beta_{2} $ of the observer are adjusted appropriately, the system speed estimation value $\ \ z_{1} $  and the total disturbance estimation value $\ z_{2} $ can be obtained through the state observer. It can be seen from equation (23) that if the system $\ d_{\omega} $ is observed and compensated, the PMSM speed control system can be approximated as a first-order integral system.

The linear state error feedback control law is adopted for the input $\ u_{0} $, so the control law based on disturbance compensation controller can be obtained as
\begin{align}
\label{Mechanica21}
\left\{ \begin{array}{l}
{u_0} = {k_p}(\omega _{}^* - {z_1})\\
{i_q}^* = {u_0} - k{z_2}
\end{array} \right.
\end{align}
where $k = \frac{{2J}}{{3p{\psi _f}}}$,$\ k_{p} $ is the magnification factor,$\ \omega _{{}}^*$ is the reference speed.
\section{Simulation and analysis}
In order to test the effectiveness of the control strategy proposed in Fig.3, some simulation results are given on the Matlab/simulink. The following three groups of simulation results are carried out. The first is to show the steady-state performance of IM MPCC for PMSM  by comparing with conventional MPCC. The second is to verify that the proposed strategy can improve the dynamic response of the system by comparing with proposed IM MPCC. The last is to reflect the high-order harmonic reduction effect of the current control strategy by total harmonic distortion (THD). The specific simulation parameters of the surface-mounted permanent magnet synchronous motor are shown in Table 1.
\begin{table}[htbp]
\caption{Parameter Values of PMSM}
\begin{center}
\begin{tabular}{l|l|l}
 \hline
  {\bf Descriptions} & {\bf Parameters} & {\bf Nominal Values} \\ \hline
   DC link Voltage  & $V_{dc}$ & 311 (V) \\
   Stator Resistance & $R_s$ & 1.3 ($\Omega$) \\
   Stator Inductance & $L_s$ & 0.0085 (H) \\
   Flux & $\psi_f$ & 0.175 (Wb) \\
   Pole Pairs & $P$ & 4.0  \\
   Inertia & $J$ & 0.008 (Kg$\cdot$ m$^{2}$) \\\hline

\end{tabular}
\label{tab1}
\end{center}
\end{table}

\subsection{Steady-state Performance}
In order to test the control strategy has improved the steady-state performance of the system, the single-step model predictive current control (MPCC) strategy is compared with the proposed improved multi-step model predictive current control (IM MPCC) strategy in this part. The simulation results are shown in Fig 4. Except for the controller differences mentioned above, these two simulations have exactly the same structure and parameters. It can be seen that ripple of three-phase current (take A phase current $\ i_a$ as an example), torque ($\ T_e$) and speed ($\ \omega_{}$) are all reduced.


\begin{figure}[]
\def\siz{0.65}
\def\sizet{0.55}
\def\size{1.0}
\centering
\psfrag{i}[c][c][\siz]{$\ i_a$[A]}
\psfrag{c}[c][c][\siz]{$\ i_a$[A]}
\psfrag{e}[c][c][\siz]{$\ T_e$[N$\cdot$m]}
\psfrag{s}[c][c][\siz]{$\ \omega_{}$[rpm]}
\psfrag{T}[c][c][\siz]{Time[s]}
\psfrag{t}[c][c][\siz]{Time[s]}
\psfrag{t}[c][c][\siz]{Time[s]}
\psfrag{1}[c][c][\siz]{1}
\psfrag{0.5}[c][c][\siz]{0.5}
\psfrag{0}[c][c][\siz]{0}
\psfrag{-0.5}[c][c][\siz]{-0.5}

\psfrag{0.02}[c][c][\siz]{0.02}
\psfrag{0.03}[c][c][\siz]{0.03}
\psfrag{0.04}[c][c][\siz]{0.04}
\psfrag{0.05}[c][c][\siz]{0.05}
\psfrag{0.06}[c][c][\siz]{0.06}
\psfrag{0.07}[c][c][\siz]{0.07}
\psfrag{0.08}[c][c][\siz]{0.08}
\psfrag{0.09}[c][c][\siz]{0.09}
\psfrag{0.1}[c][c][\siz]{0.1}

\psfrag{0.115}[c][c][\siz]{}
\psfrag{0.12}[c][c][\siz]{0.12}
\psfrag{0.125}[c][c][\siz]{}
\psfrag{0.13}[c][c][\siz]{0.13}
\psfrag{0.135}[c][c][\siz]{}
\psfrag{0.14}[c][c][\siz]{0.14}
\psfrag{0.145}[c][c][\siz]{}
\psfrag{0.15}[c][c][\siz]{0.15}

\psfrag{6}[c][c][\siz]{6}
\psfrag{-6}[c][c][\siz]{}
\psfrag{5.5}[c][c][\siz]{5.5}
\psfrag{6}[c][c][\siz]{6}
\psfrag{4.5}[c][c][\siz]{4.5}
\psfrag{4}[c][c][\siz]{4}
\psfrag{-4}[c][c][\siz]{-4}
\psfrag{0.16}[c][c][\siz]{0.16}
\psfrag{0.17}[c][c][\siz]{0.17}
\psfrag{0.18}[c][c][\siz]{0.18}

\psfrag{200}[c][c][\siz]{200}
\psfrag{400}[c][c][\siz]{400}
\psfrag{600}[c][c][\siz]{600}
\psfrag{800}[c][c][\siz]{800}
\psfrag{1000}[c][c][\siz]{1000}

\psfrag{0.02}[c][c][\siz]{0.02}
\psfrag{0.8}[c][c][\siz]{0.8}
\psfrag{5}[c][c][\siz]{5}
\psfrag{-5}[c][c][\siz]{-5}
\psfrag{-2}[c][c][\siz]{-2}
\psfrag{0.11}[c][c][\siz]{0.11}
\psfrag{0.01}{0.01}
\psfrag{-0.8}[c][c][\siz]{-0.8}
\psfrag{1000.2}[c][c][\siz]{1000.2}
\psfrag{999.8}[c][c][\siz]{999.8}
\psfrag{999.6}[c][c][\siz]{}
\psfrag{0.075}[c][c][\siz]{0.075}
\psfrag{0.085}[c][c][\siz]{0.085}
\psfrag{0.095}[c][c][\siz]{0.095}
\psfrag{2}[c][c][\siz]{2}
\psfrag{8}[c][c][\siz]{8}
\psfrag{MPCC}[c][c][\sizet]{MPCC}
\psfrag{IM MPCC}[c][c][\sizet]{IM MPCC}
\psfrag{Reference}[c][c][\sizet]{Reference}
\includegraphics[height=3cm,width=9.3cm]{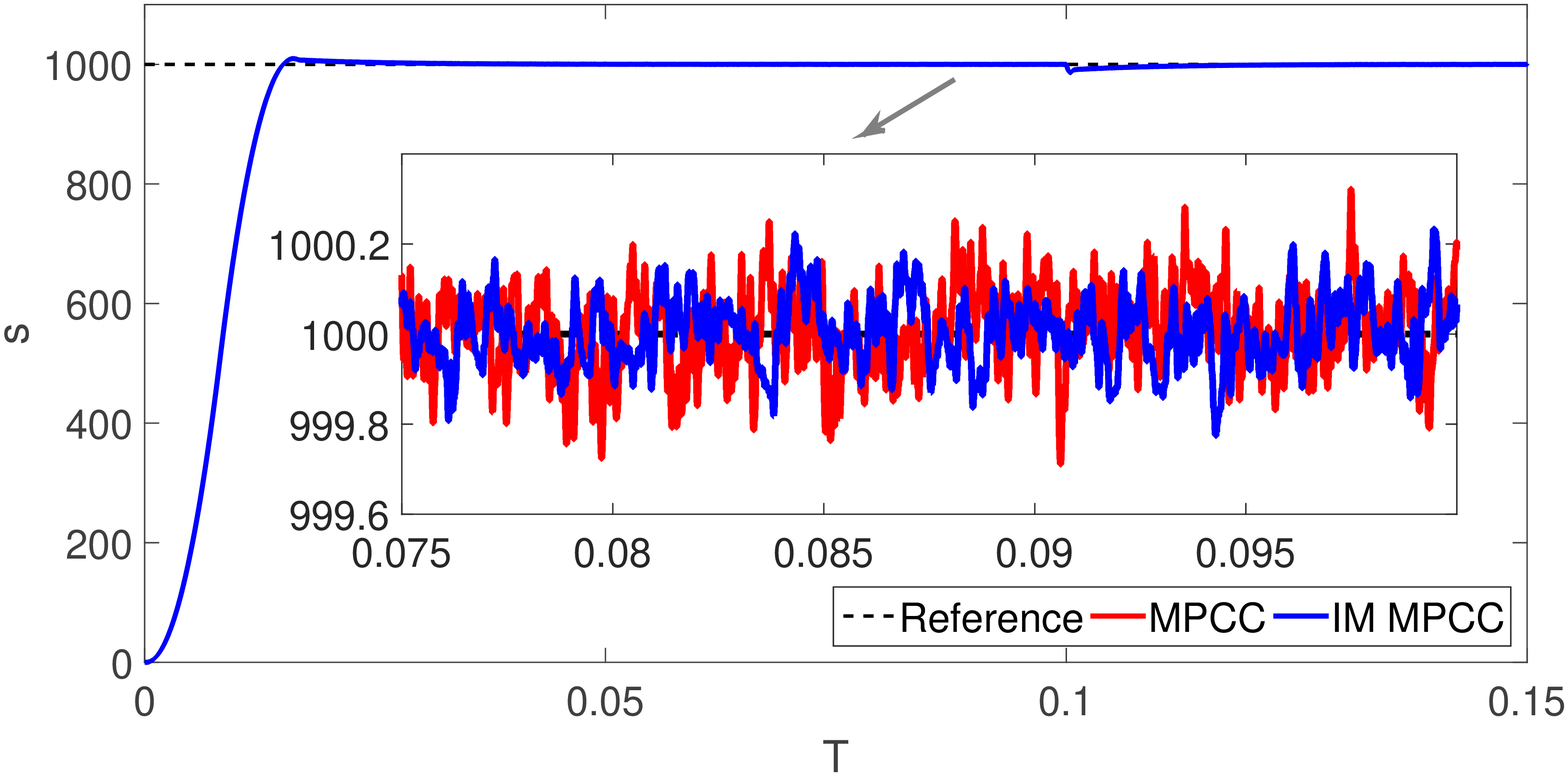}
\begin{center}(a)
\end{center}
\includegraphics[height=2.8cm,width=9.3cm]{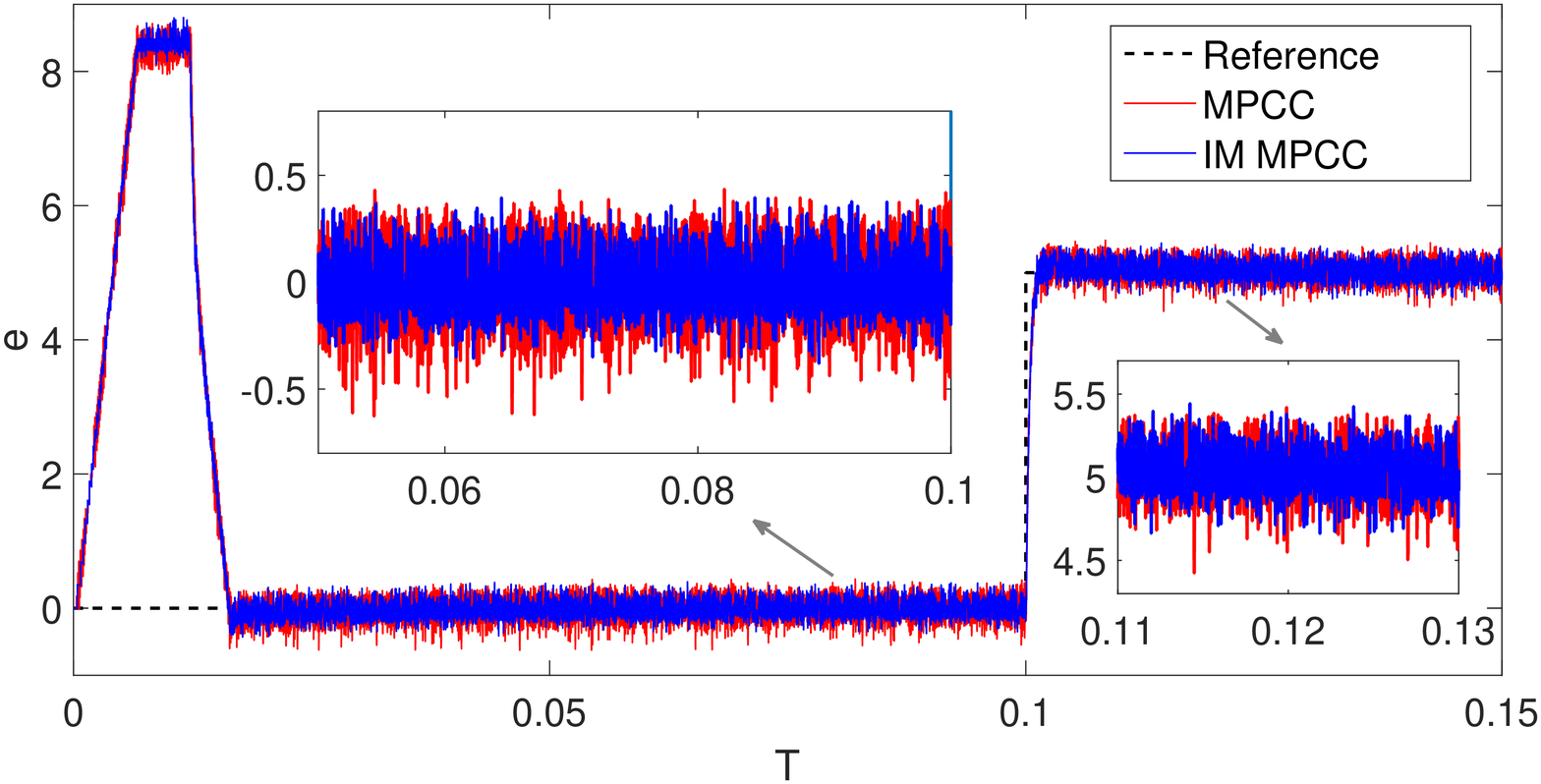}
\begin{center}(b)
\end{center}
\includegraphics[height=3cm,width=9.3cm]{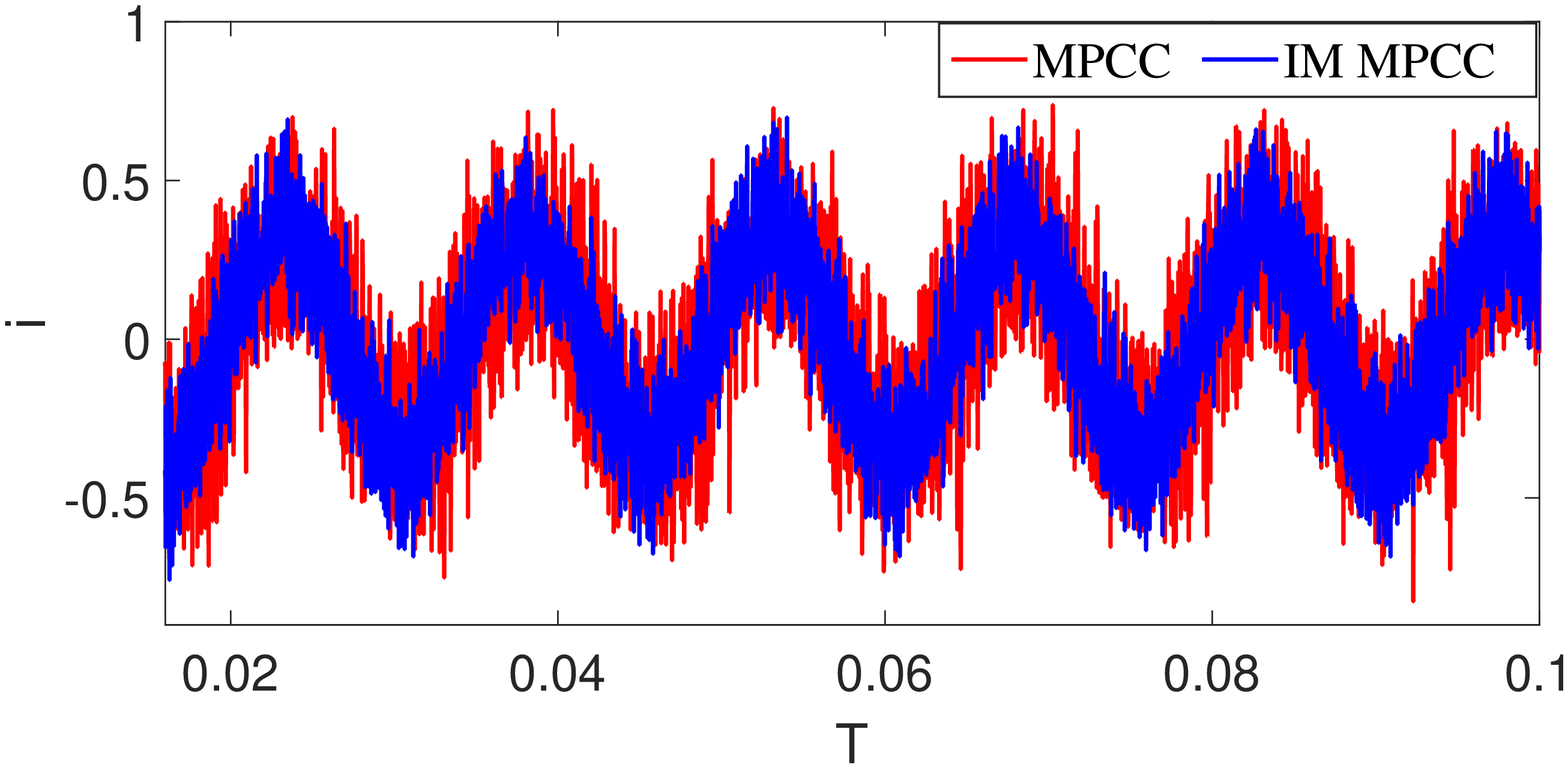}
\begin{center}(c)
\end{center}
\includegraphics[height=3cm,width=9.3cm]{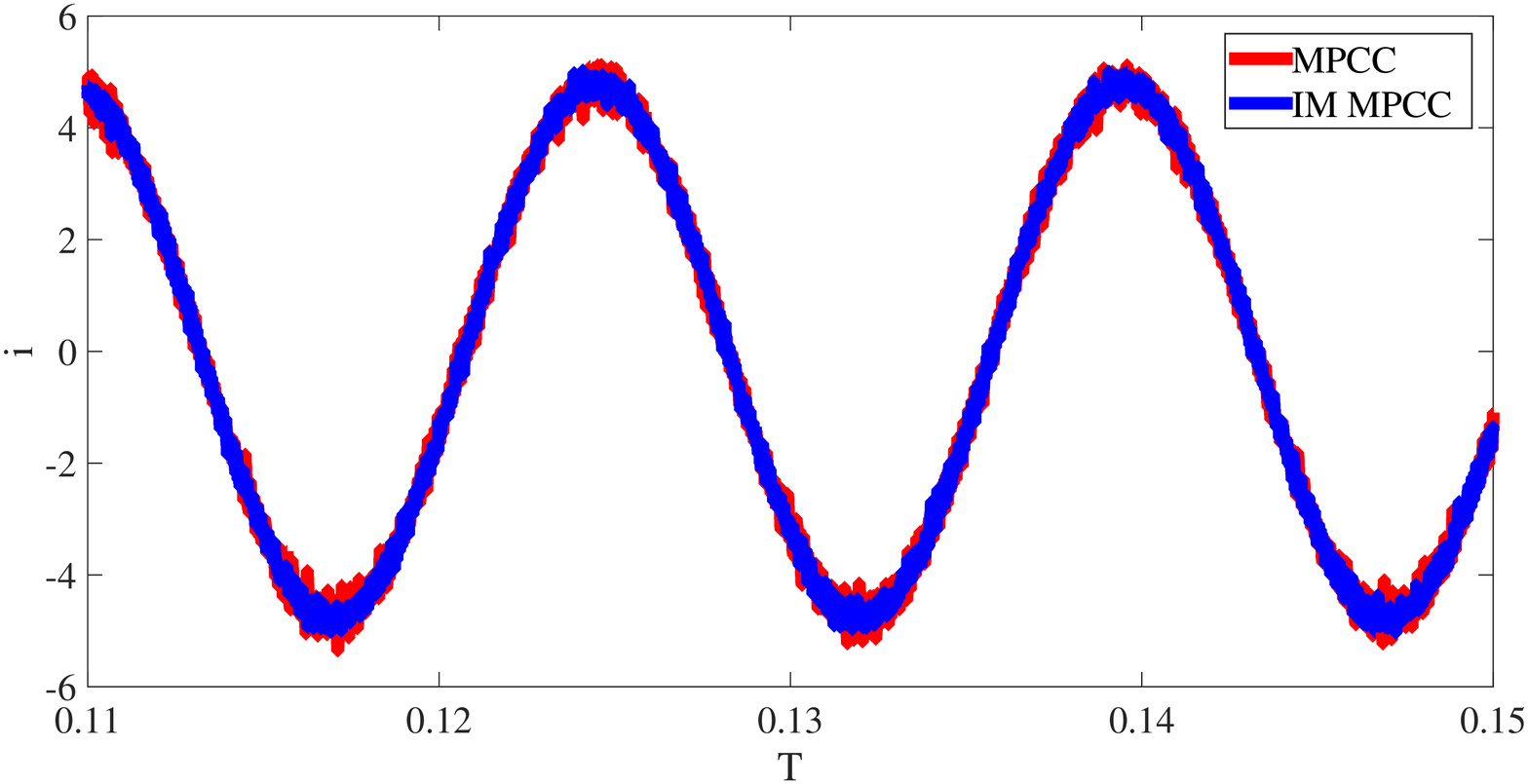}
\begin{center}(d)
\end{center}
\caption{ (Simulation). Steady-state performance comparison between conventional MPCC and IM MPCC: (a) Speed; (b) Torque; (c) Phase-current under zero load; and (d) Phase current under  5N$\cdot$m load.}
\end{figure}

\subsection{Dynamic Response}
In order to fairly test the dynamic response of the proposed control strategy, the improved multi-step model predictive control strategy with speed loop disturbance compensation (IM MPCC+DC) is compared with the control strategy without disturbance compensation. Except for this differences mentioned above, these two simulations have exactly the same structure and parameters. The dynamic response of the system results from the sudden change of load disturbance. The load disturbance is suddenly increased , and then suddenly reduced to 0 (load torque from 0 to 5 N$\cdot$m, and then jumps back to 0), the effect is shown in Fig 5. It can be seen that dynamic response of IM MPCC with speed loop DC is better than the strategy without DC.


\begin{figure}[]
\def\siz{0.65}
\def\size{1.0}
\def\sizet{0.55}
\centering
\psfrag{TE}[c][c][\siz]{$\ T_e$[N$\cdot$m]}
\psfrag{s}[c][c][\siz]{$\ \omega_{}$[rpm]}
\psfrag{IM MPCC}[c][c][\sizet]{IM MPCC}
\psfrag{Reference}[c][c][\sizet]{Reference}
\psfrag{IM MPCC+DC}[c][c][\sizet]{IM MPCC+DC}
\psfrag{T}[c][c][\siz]{Time[s]}
\psfrag{t}[c][c][\siz]{Time[s]}
\psfrag{5}[c][c][\siz]{5}
\psfrag{6}[c][c][\siz]{6}
\psfrag{4}[c][c][\siz]{4}
\psfrag{3}[c][c][\siz]{3}
\psfrag{2}[c][c][\siz]{2}
\psfrag{1}[c][c][\siz]{1}
\psfrag{-1}[c][c][\siz]{-1}
\psfrag{0}[c][c][\siz]{0}

\psfrag{-5}[c][c][\siz]{-5}
\psfrag{0.11}[c][c][\siz]{0.11}
\psfrag{0.12}[c][c][\siz]{0.12}
\psfrag{0.055}[c][c][\siz]{}
\psfrag{1050}[c][c][\siz]{1050}
\psfrag{1000}[c][c][\siz]{1000}
\psfrag{950}[c][c][\siz]{950}
\psfrag{0.06}[c][c][\siz]{0.06}
\psfrag{0.065}[c][c][\siz]{}
\psfrag{0.07}[c][c][\siz]{0.07}
\psfrag{0.075}[c][c][\siz]{}
\psfrag{0.08}[c][c][\siz]{0.08}
\psfrag{0.085}[c][c][\siz]{}
\psfrag{0.09}[c][c][\siz]{0.09}
\psfrag{0.095}[c][c][\siz]{}
\psfrag{0.1}[c][c][\siz]{0.1}
\psfrag{0.062}[c][c][\siz]{0.062}
\psfrag{0.064}[c][c][\siz]{0.064}
\psfrag{0.086}[c][c][\siz]{0.086}
\psfrag{0.088}[c][c][\siz]{0.088}
\psfrag{0.09}[c][c][\siz]{0.09}

\includegraphics[height=3cm,width=9cm]{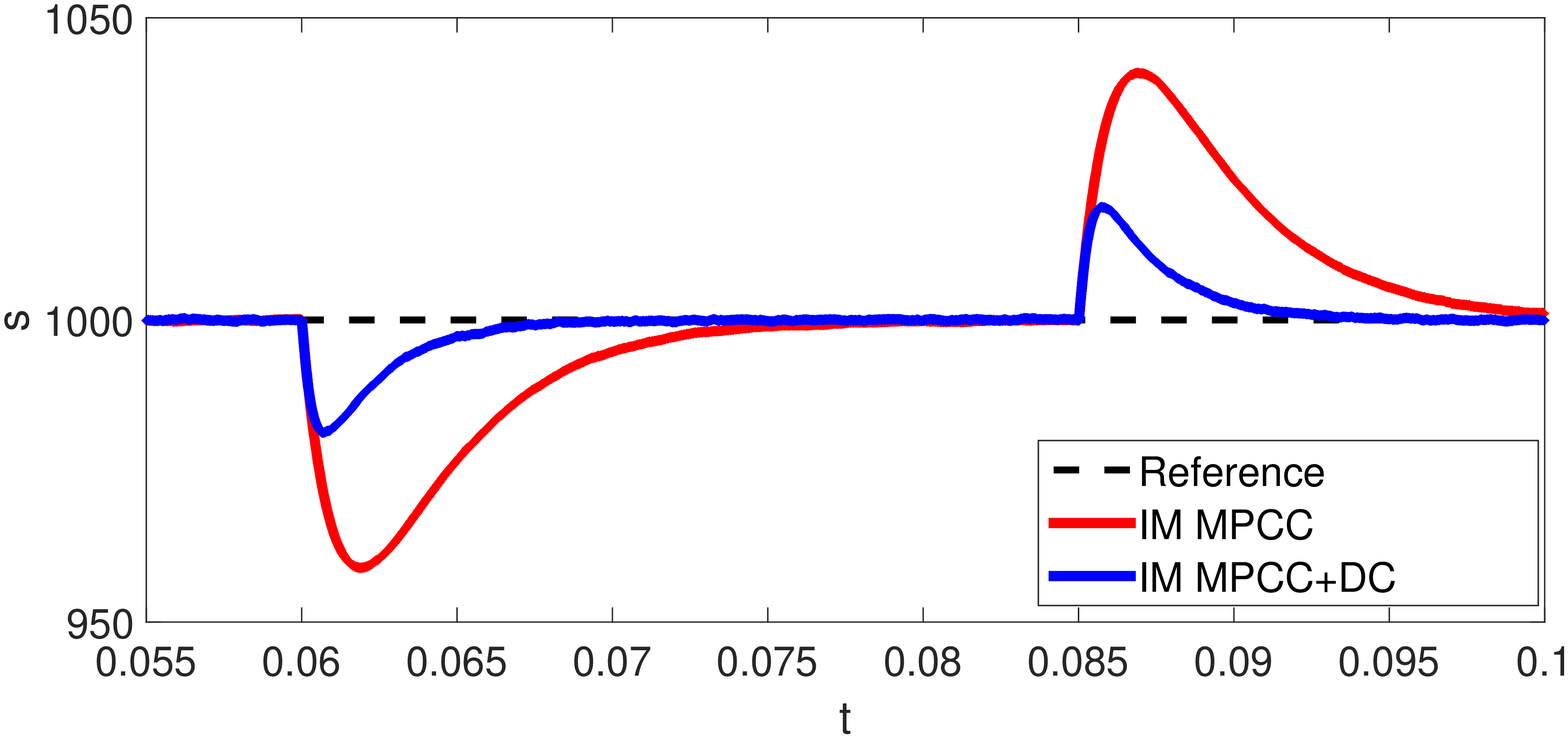}
\begin{center}(a)
\end{center}
\includegraphics[height=3cm,width=9cm]{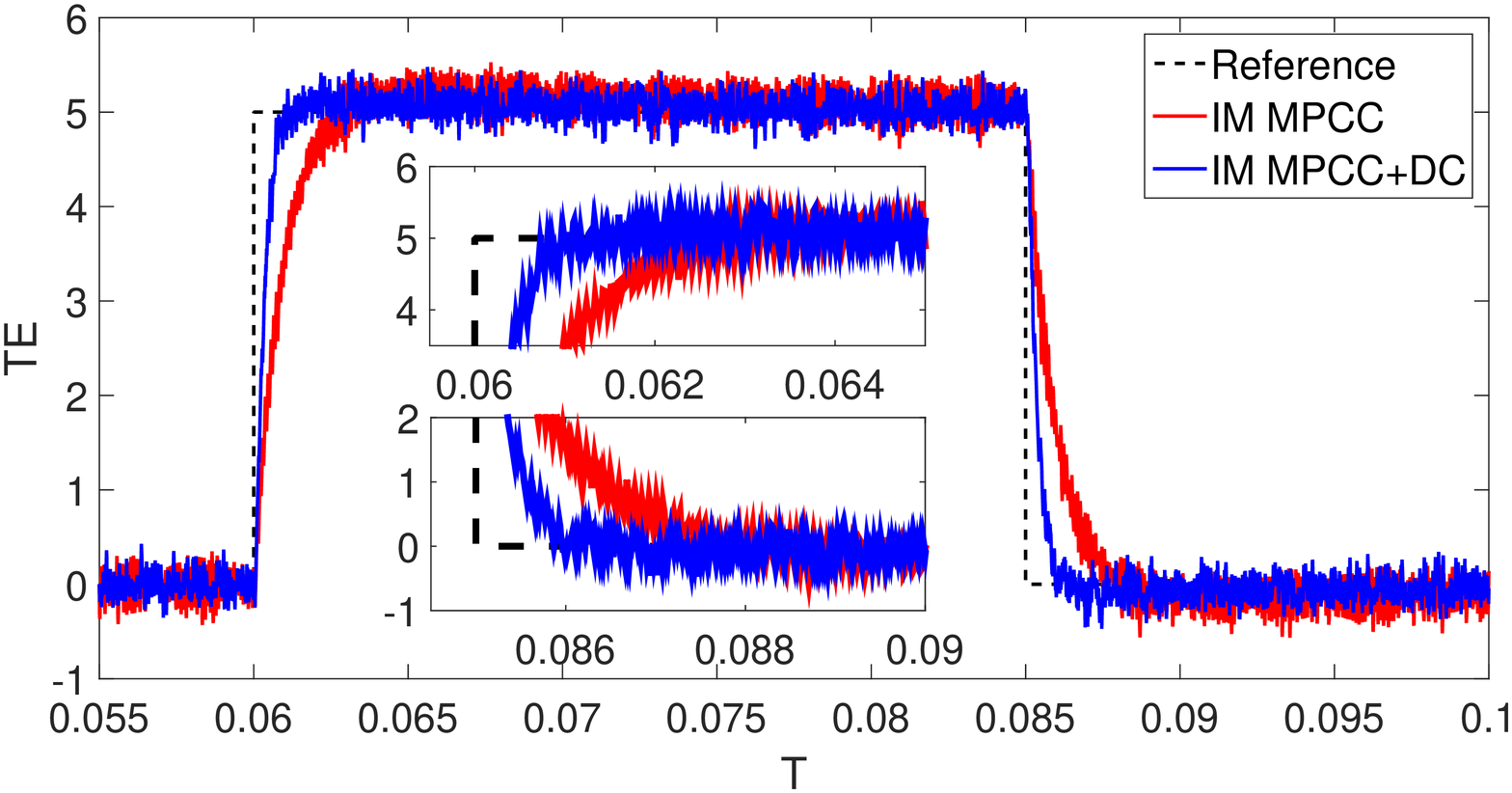}
\begin{center}(b)
\end{center}
\caption{(Simulation). Response curves of the IM MPCC strategy with and without disturbance compensation: (a) Speed; and (b) Torque.}
\label{fig:15001rpm}
\end{figure}
\subsection{THD }
By comparing the high-frequency component with the basic component of the three-phase current, we can see that the three-phase current THD data of the three strategies are shown in Table 2. It can be observed that compared with the initial single-step prediction measurement, the THD of the three-phase current of the proposed control strategy is reduced by 27.18$\ \rm{\% }$ on average.

\begin{table}[htbp]
\caption{Three-phase Current THD Comparison (Simulation)}
\begin{center}
\begin{tabular}{|c|c|c|c|}
\hline
\textbf{Control}&\multicolumn{3}{|c|}{\textbf{Three-phase Current THD}} \\
\cline{2-4}
\textbf{Strategy} & \textbf{\textit{$\ i_a$}}& \textbf{\textit{$\ i_b$}}& \textbf{\textit{$\ i_c$}} \\
\hline
MPCC& 1.034$\ \rm{\% }$& 1.046$\ \rm{\% }$& 1.040 $\ \rm{\% }$ \\
IM MPCC& 0.773$\ \rm{\% }$& 0.796$\ \rm{\% }$& 0.792$\ \rm{\% }$ \\
IM MPCC+DC& 0.737$\ \rm{\% }$& 0.776$\ \rm{\% }$& 0.759$\ \rm{\% }$ \\
\hline
\end{tabular}
\label{tab1}
\end{center}
\end{table}
\section{Experiment and analysis}

To further verify the effectiveness of the proposed control strategy, an experiment platform is set up, which is shown by Fig. 6. The experimental platform consists of a dSPACE controller board, a three-phase inverter, a DC link power supply, current sensors, a PMSM and a load motor. The dSPACE 1103 controller board used in this experimental test setup integrates a core
processor based on TMS320F240 DSP. The dSPACE mainly implements signal acquisition, signal processing and controller calculation for output, and the minimum reliable current
measurement time is set to 30 $\mu s$. The three-phase inverter is built by a 3 kW FSBB30CH60C Smart Power Module (SPM) with a rated current of 10 A. To ensure the accuracy of the
feedback current, LEM LAH25-NP is employed as the current sensor. The maximum voltage of the DC link power supply is 300 V.

\begin{figure}[htbp]
\def\siz{0.65}
\def\sizet{0.55}
\def\size{1.0}
\begin{center}
\psfrag{u3}{t3}
\psfrag{010}{8}
\includegraphics[height=5cm,width=7cm]{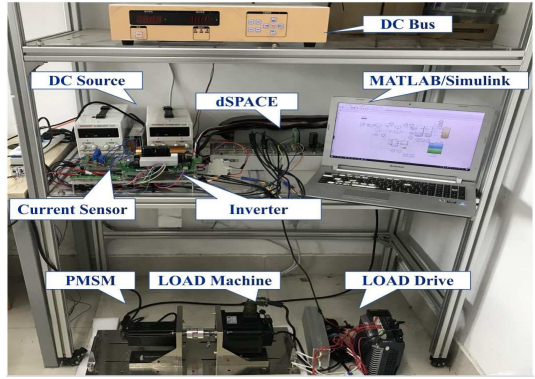}
\end{center}
\caption{Experimental platform.}
\label{1}
\end{figure}

\subsection{Steady-state Performance}

In order to test the performance of the two control strategies fairly, it is necessary to ensure that the parameters in the prediction model are consistent. The parameters of the two control policies are selected as follows. The traditional PI control is aodpted in the speed loop, where the parameters are designed as $K_p = 0.003$ and $K_i = 0.001$. The reference speed is set as 1000 rpm. The experimental results of the MPCC and IMMPCC strategies in steady state are shown in Fig. 7. The maximum speed and torque offsets of the IM MPCC strategy were 2.31 rpm and 0.19 N$\cdot$m, respectively, which improved the control performance as compared with these of 4.35 rpm and 0.28 N$\cdot$m of the MPCC strategy.

\begin{figure}[htbp]
\def\siz{0.65}
\def\sizet{0.55}
\def\size{1.0}
\begin{center}
\psfrag{u3}{t3}
\psfrag{010}{8}
\psfrag{0}[c][c][\siz]{0}
\psfrag{0}[c][c][\size]{0}\psfrag{2}[c][c][\size]{2}
\psfrag{4}[c][c][\size]{4}\psfrag{6}[c][c][\size]{6}
\psfrag{8}[c][c][\size]{8}\psfrag{10}[c][c][\size]{10}
\psfrag{12}[c][c][\size]{12}\psfrag{500}[c][c][\size]{500}
\psfrag{1000}[c][c][\size]{1000}\psfrag{1020}[c][c][\size]{1020}
\psfrag{980}[c][c][\size]{980}\psfrag{7}[c][c][\size]{7}
\psfrag{980}[c][c][\size]{980}\psfrag{8}[c][c][\size]{8}
\psfrag{4.5}[c][c][\size]{}\psfrag{6.8}[c][c][\size]{6.8}
\psfrag{5.5}[c][c][\size]{}\psfrag{5}[c][c][\size]{5}
\psfrag{666}[c][c][\size]{}
\psfrag{777}[c][c][\size]{$\omega$[rpm]}
\includegraphics[height=2.5cm,width=9cm]{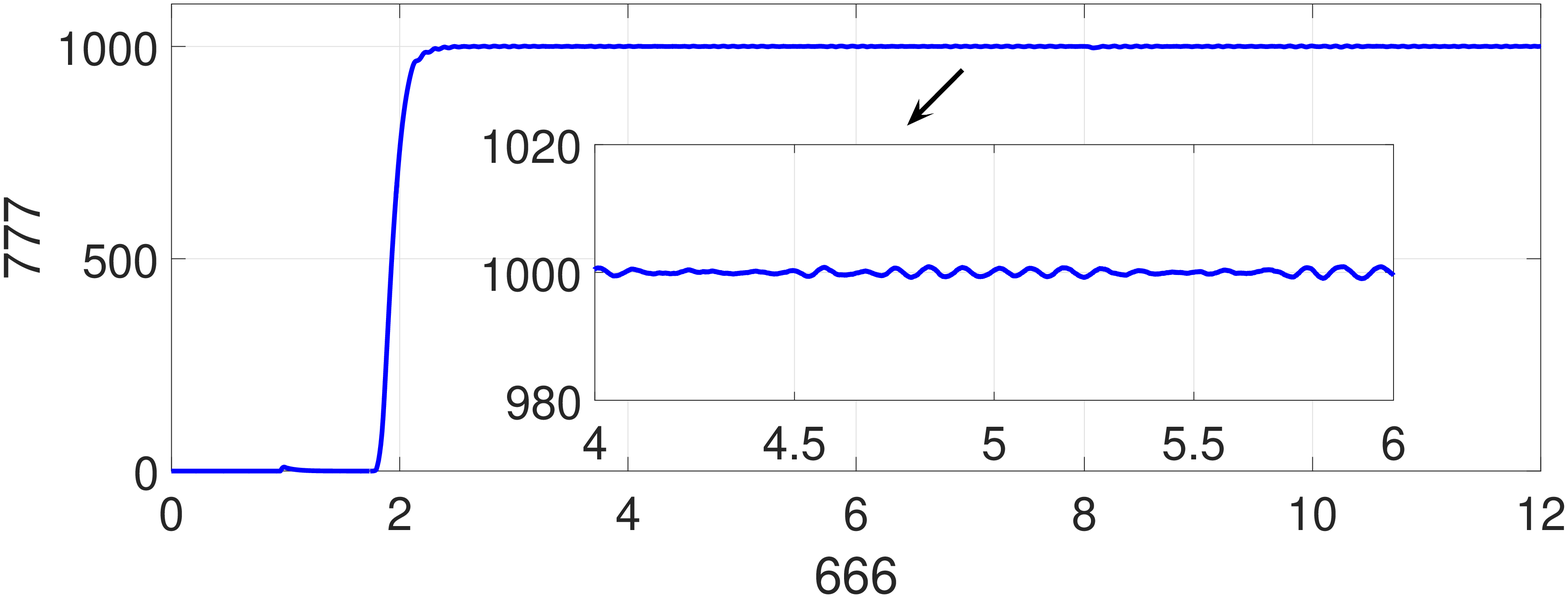}
\psfrag{9}[c][c][\size]{0}\psfrag{10}[c][c][\size]{0.1}
\psfrag{11}[c][c][\size]{0.2}\psfrag{12}[c][c][\size]{0.3}
\psfrag{13}[c][c][\size]{0.4}\psfrag{0.15}[c][c][\size]{1}
\psfrag{0.25}[c][c][\size]{2}
\psfrag{888}[c][c][\size]{$T_e$[N$\cdot$m]}
\includegraphics[height=2.5cm,width=9cm]{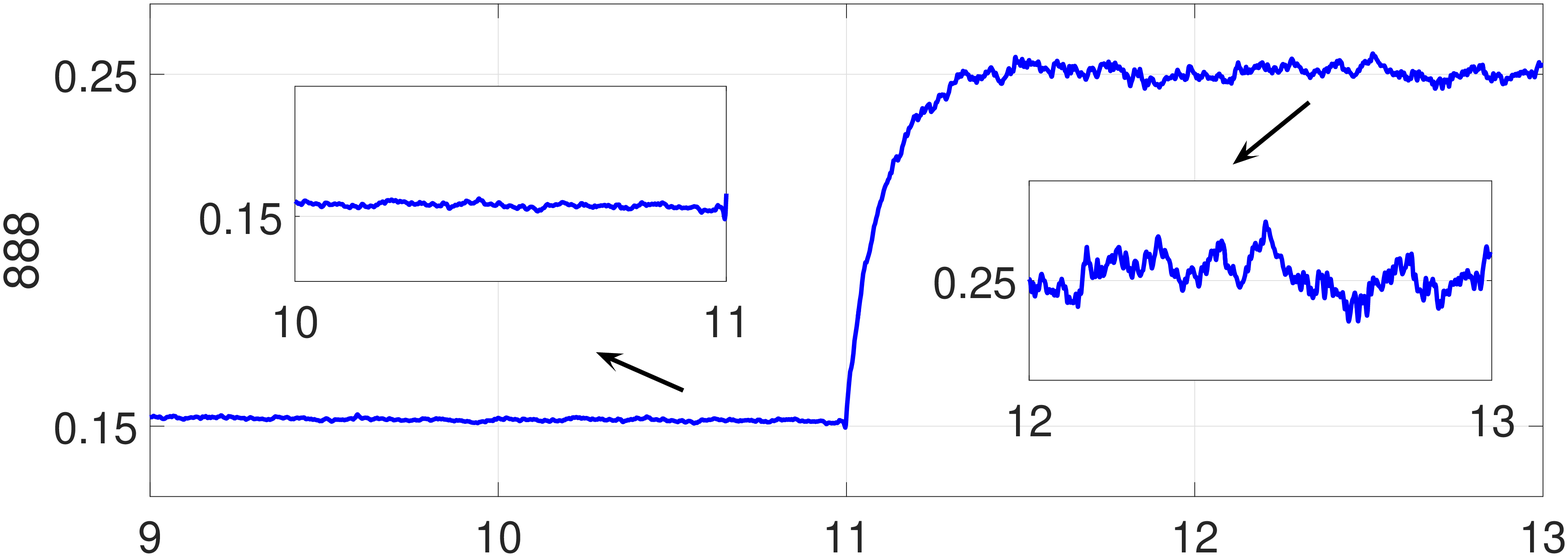}
\psfrag{10.9}[c][c][\size]{0}\psfrag{10.95}[c][c][\size]{0.05}
\psfrag{11}[c][c][\size]{0.1}\psfrag{0.2}[c][c][\size]{0.2}
\psfrag{0}[c][c][\size]{0}\psfrag{-0.2}[c][c][\size]{-0.2}
\psfrag{0.1}[c][c][\size]{0.1}
\psfrag{-0.1}[c][c][\size]{-0.1}
\psfrag{999}[c][c][\size]{$i_{abc}$[A]}
\psfrag{d1}[c][c][\size]{$\ i_{a}$}
\psfrag{d2}[c][c][\size]{$\ i_{b}$}
\psfrag{d3}[c][c][\size]{$\ i_{c}$}
\includegraphics[height=2.5cm,width=9cm]{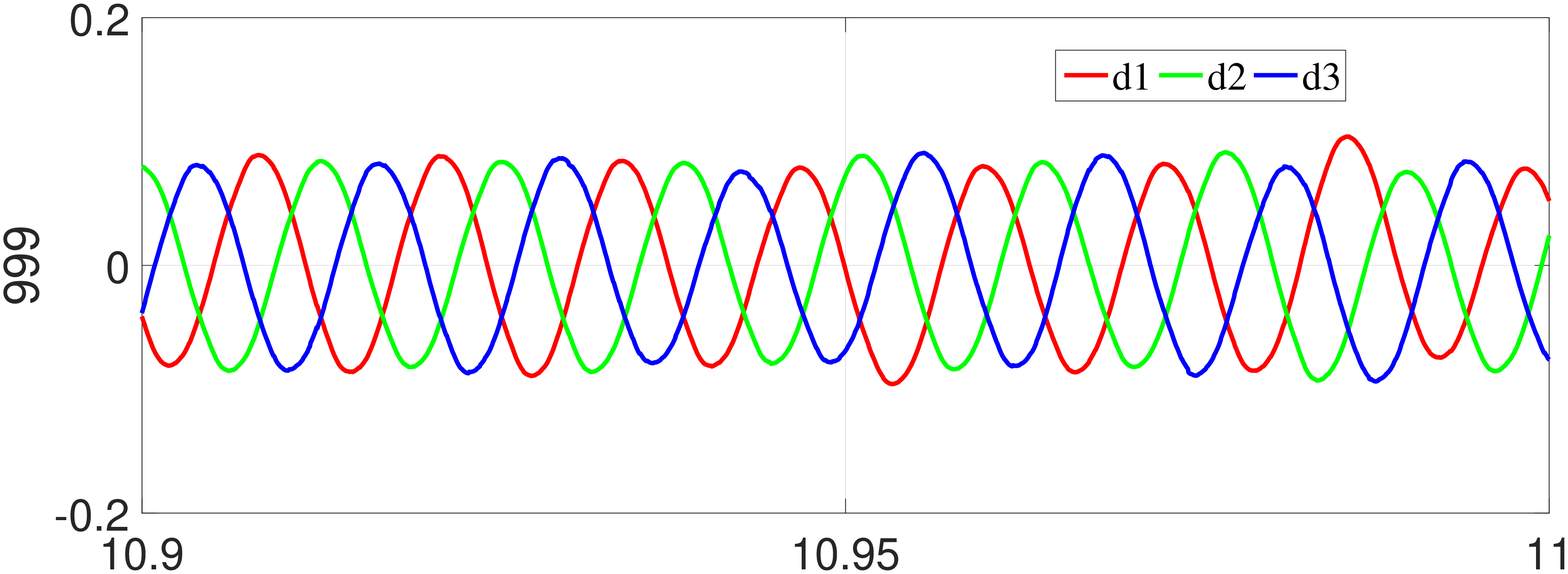}
\psfrag{d1}[c][c][\size]{$\ i_{a}$}
\psfrag{d2}[c][c][\size]{$\ i_{b}$}
\psfrag{d3}[c][c][\size]{$\ i_{c}$}
\psfrag{14}[c][c][\size]{0}\psfrag{14.05}[c][c][\size]{0.05}
\psfrag{14.1}[c][c][\size]{0.1}\psfrag{0.2}[c][c][\size]{0.2}
\psfrag{0}[c][c][\size]{0}\psfrag{-0.2}[c][c][\size]{-0.2}
\psfrag{999}[c][c][\size]{$i_{abc}$[A]}
\psfrag{555}[c][c][\size]{Time[s]}
\includegraphics[height=2.5cm,width=9cm]{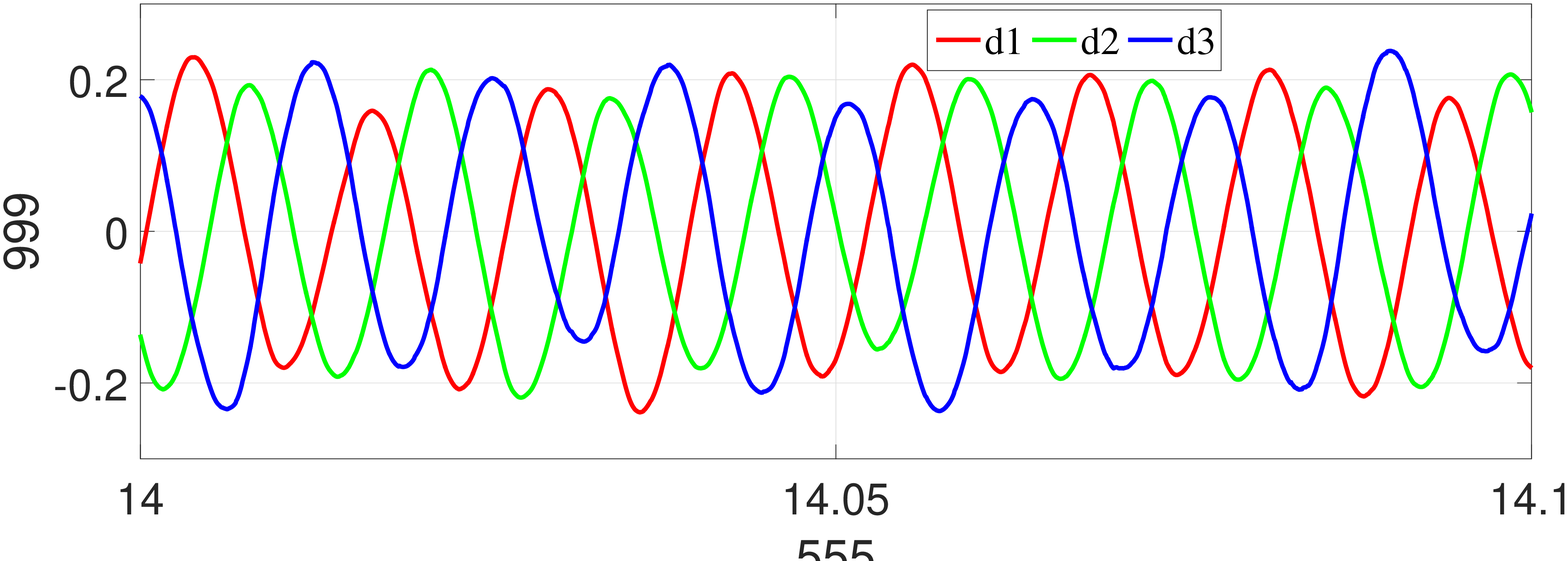}
\begin{center}(a)
\end{center}
\psfrag{0}[c][c][\siz]{0}
\psfrag{0}[c][c][\size]{0}\psfrag{2}[c][c][\size]{2}
\psfrag{4}[c][c][\size]{4}\psfrag{6}[c][c][\size]{6}
\psfrag{8}[c][c][\size]{8}\psfrag{10}[c][c][\size]{10}
\psfrag{12}[c][c][\size]{12}\psfrag{500}[c][c][\size]{500}
\psfrag{1000}[c][c][\size]{1000}\psfrag{1030}[c][c][\size]{1020}
\psfrag{970}[c][c][\size]{980}\psfrag{7}[c][c][\size]{7}
\psfrag{980}[c][c][\size]{980}\psfrag{8}[c][c][\size]{8}
\psfrag{6.5}[c][c][\size]{}\psfrag{6.8}[c][c][\size]{6.8}
\psfrag{5.5}[c][c][\size]{}\psfrag{5}[c][c][\size]{5}
\psfrag{666}[c][c][\size]{}
\psfrag{777}[c][c][\size]{$\omega$[rpm]}
\includegraphics[height=2.5cm,width=9cm]{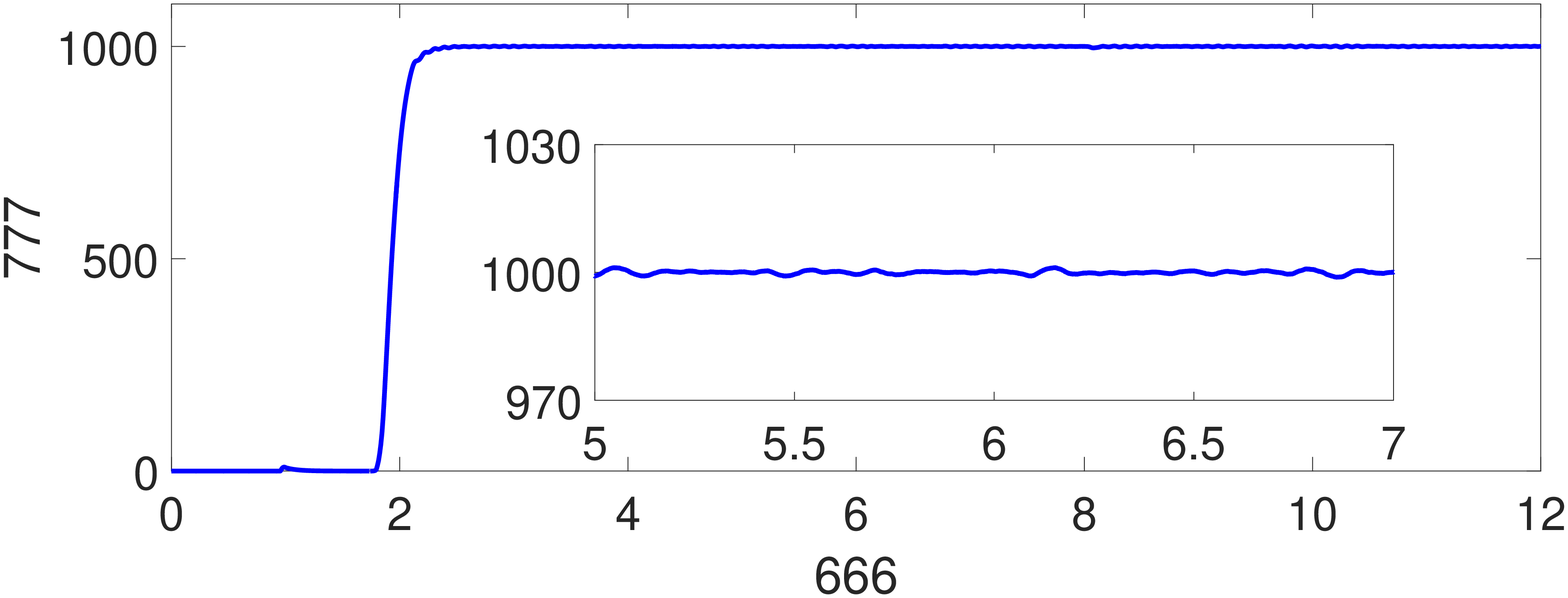}
\psfrag{9}[c][c][\size]{0}\psfrag{10}[c][c][\size]{0.1}
\psfrag{11}[c][c][\size]{0.2}\psfrag{12}[c][c][\size]{0.3}
\psfrag{13}[c][c][\size]{0.4}\psfrag{0.2}[c][c][\size]{1}
\psfrag{0.25}[c][c][\size]{2}
\psfrag{888}[c][c][\size]{$T_e$[N$\cdot$m]}
\includegraphics[height=2.5cm,width=9cm]{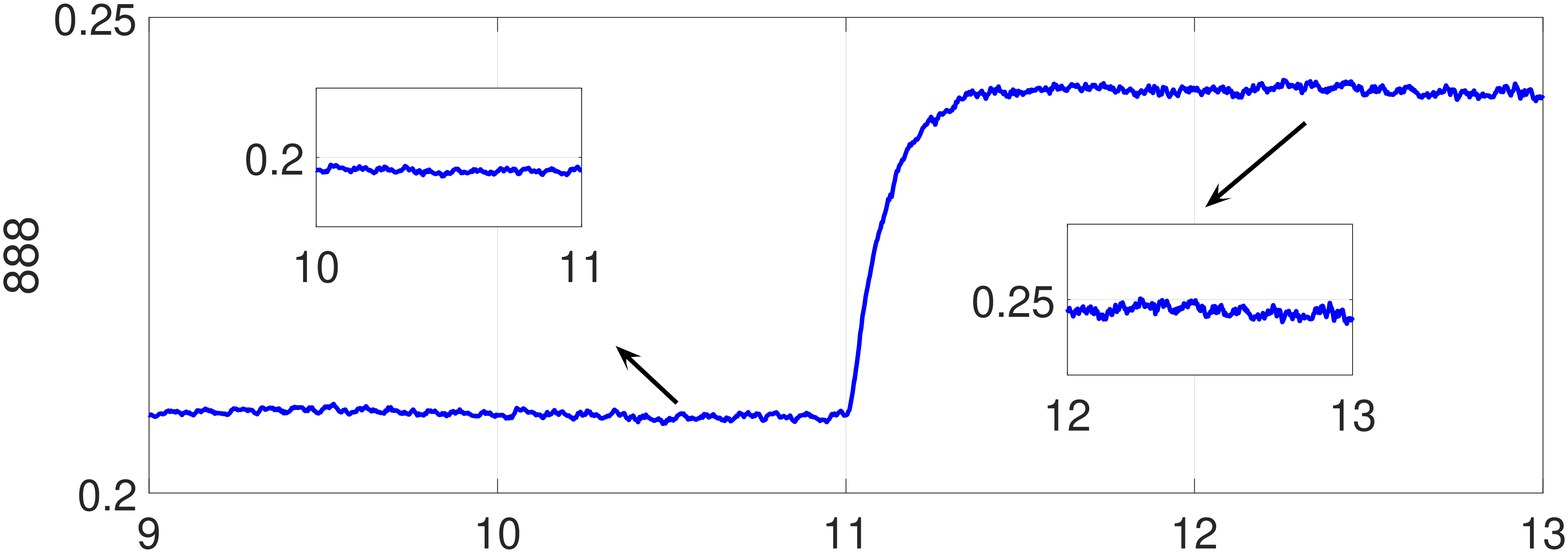}
\psfrag{12.8}[c][c][\size]{0}\psfrag{12.85}[c][c][\size]{0.05}
\psfrag{12.9}[c][c][\size]{0.1}\psfrag{0.2}[c][c][\size]{0.2}
\psfrag{0}[c][c][\size]{0}\psfrag{-0.2}[c][c][\size]{-0.2}
\psfrag{999}[c][c][\size]{$i_{abc}$[A]}
\psfrag{d1}[c][c][\size]{$\ i_{a}$}
\psfrag{d2}[c][c][\size]{$\ i_{b}$}
\psfrag{d3}[c][c][\size]{$\ i_{c}$}
\includegraphics[height=2.5cm,width=9cm]{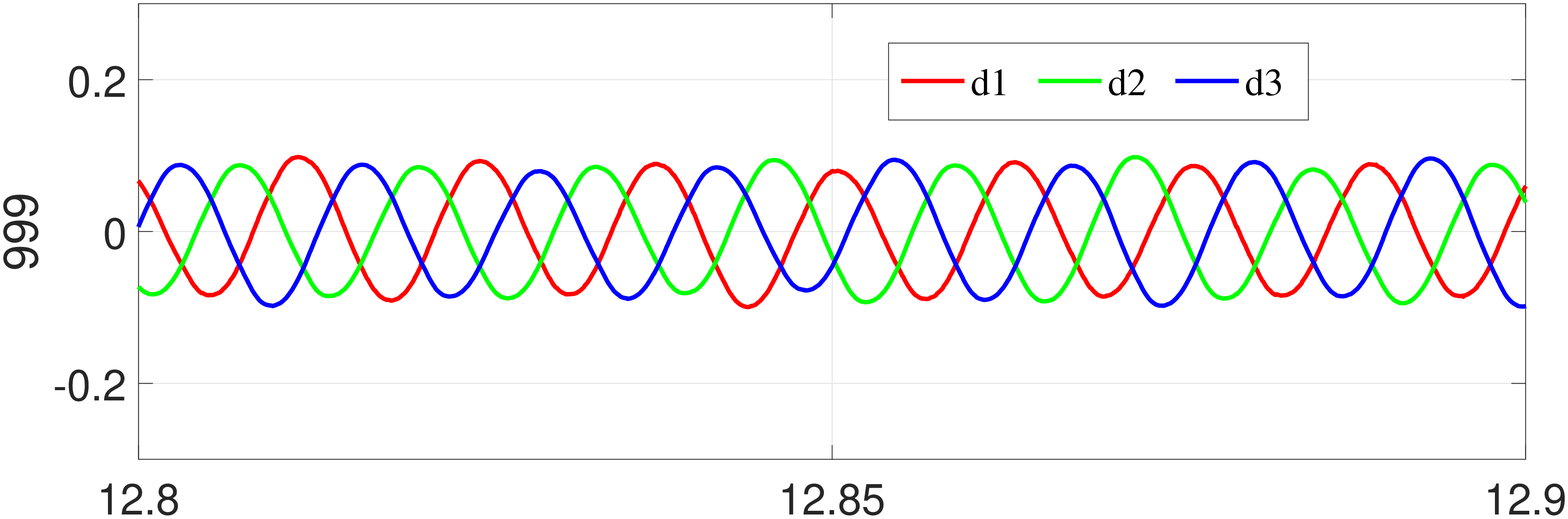}
\psfrag{19}[c][c][\size]{0}\psfrag{19.05}[c][c][\size]{0.05}
\psfrag{19.1}[c][c][\size]{0.1}\psfrag{0.1}[c][c][\size]{0.2}
\psfrag{0}[c][c][\size]{0}\psfrag{-0.1}[c][c][\size]{-0.2}
\psfrag{999}[c][c][\size]{$i_{abc}$[A]}
\psfrag{d1}[c][c][\size]{$\ i_{a}$}
\psfrag{d2}[c][c][\size]{$\ i_{b}$}
\psfrag{d3}[c][c][\size]{$\ i_{c}$}
\psfrag{555}[c][c][\size]{Time[s]}
\includegraphics[height=2.5cm,width=9cm]{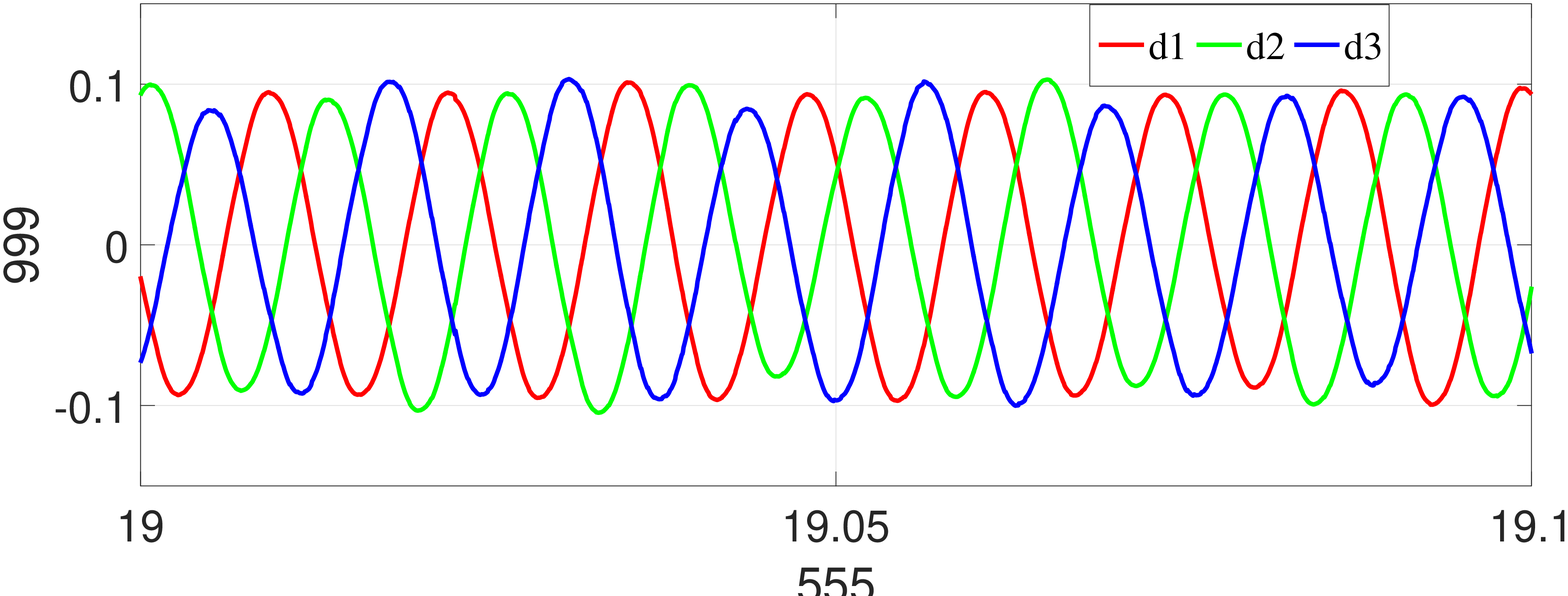}
\begin{center}(b)
\end{center}
\end{center}
\caption{(Experiment). Response curves for steady-state performance evaluation: (a) the traditional  MPCC; and (b) the proposed IM MPCC.}
\label{1}
\end{figure}

\subsection{Dynamic Response}

Fig. 8 shows the experimental results of IM MPCC strategy when the system is subjected to a load torque disturbance. The load torque was suddenly increased to 2 N$\cdot$m, and the reference speed was set at 1000 rpm. Two sets of experimental results that  with or without disturbance compensation (DC) mechanism based on ESO are carried out for comparisons. The parameters of ESO are designed as $\beta_1 = 1200$ and $\beta_2 = 4000$. In order to quantitatively analyze the experimental results, the index parameter reflecting the dynamic performance was given in Table 3. The experimental results show that the proposed IM MPCC strategy with disturbance compensation mechanism has much faster dynamic response and disturbance rejection ability.

\begin{figure}[htbp]
\def\siz{0.65}
\def\sizet{0.55}
\def\size{1.0}
\begin{center}
\psfrag{u3}{t3}
\psfrag{010}{8}
\psfrag{0}[c][c][\size]{0}
\psfrag{960}[c][c][\size]{980}\psfrag{1000}[c][c][\size]{1000}
\psfrag{1040}[c][c][\size]{1020}\psfrag{6.5}[c][c][\size]{0}
\psfrag{7.5}[c][c][\size]{0.1}\psfrag{8.5}[c][c][\size]{0.2}
\psfrag{9.5}[c][c][\size]{0.3}\psfrag{10}[c][c][\size]{0}
\psfrag{11}[c][c][\size]{0.1}\psfrag{12}[c][c][\size]{0.2}
\psfrag{13}[c][c][\size]{0.3}\psfrag{0.2}[c][c][\size]{1}
\psfrag{0.25}[c][c][\size]{2}
\psfrag{666}[c][c][\size]{}
\psfrag{777}[c][c][\size]{$\omega$[rpm]}
\includegraphics[height=3cm,width=9cm]{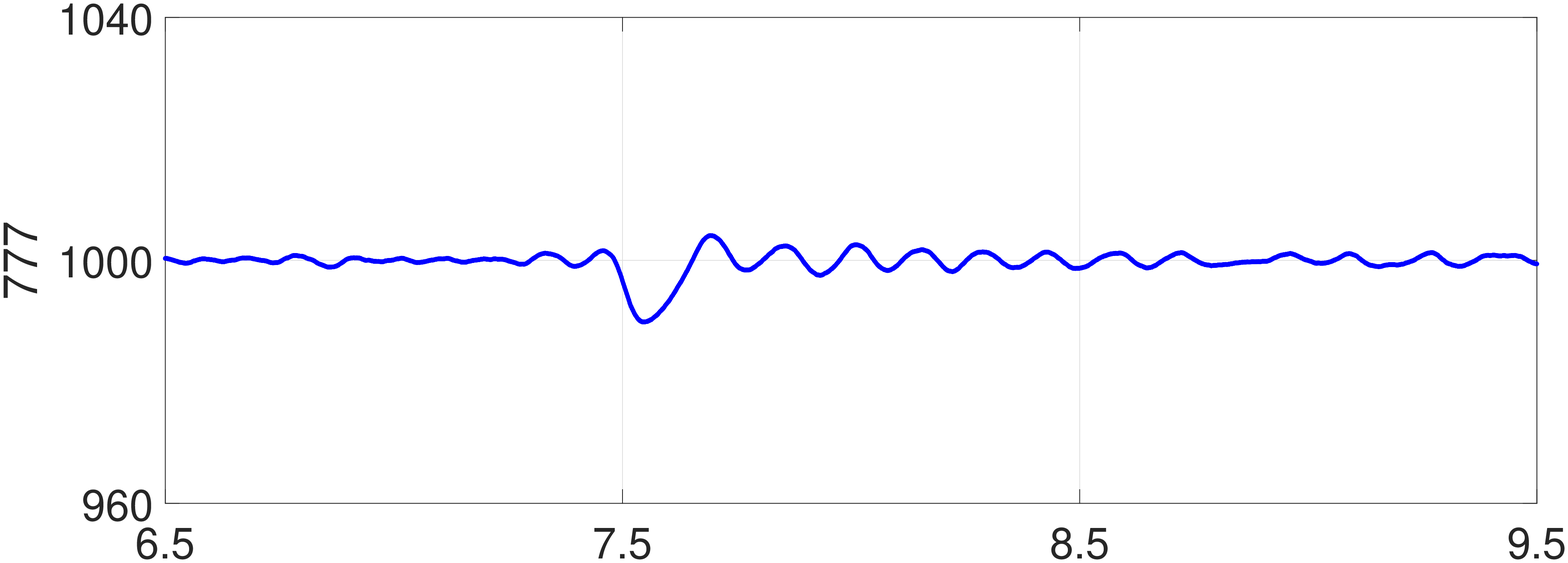}
\psfrag{888}[c][c][\size]{$T_e$[N$\cdot$m]}
\psfrag{555}[c][c][\size]{Time[s]}
\includegraphics[height=3cm,width=9cm]{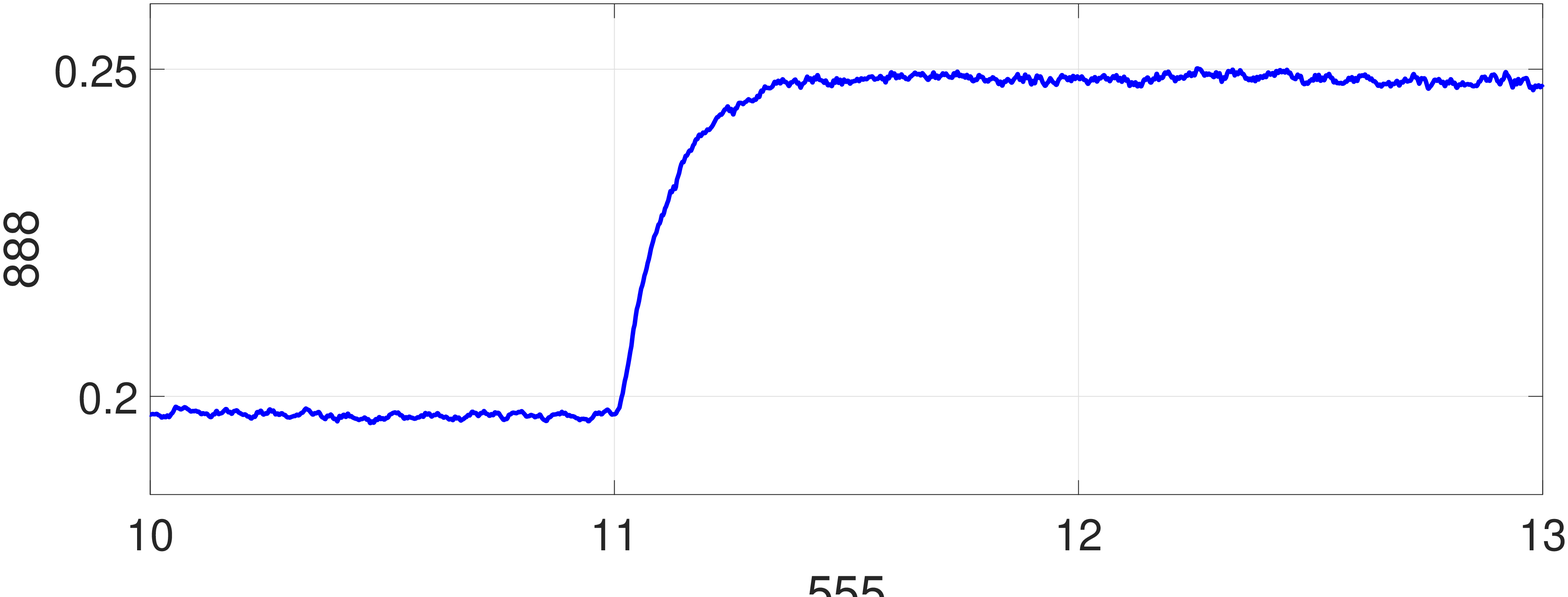}
\begin{center}(a)
\end{center}
\psfrag{0}[c][c][\size]{0}
\psfrag{960}[c][c][\size]{980}\psfrag{1000}[c][c][\size]{1000}
\psfrag{1040}[c][c][\size]{1020}\psfrag{6.5}[c][c][\size]{0}
\psfrag{7.5}[c][c][\size]{0}\psfrag{8}[c][c][\size]{0.1}
\psfrag{8.5}[c][c][\size]{0.2}\psfrag{9}[c][c][\size]{0.3}
\psfrag{10}[c][c][\size]{0}
\psfrag{11}[c][c][\size]{0.1}\psfrag{13}[c][c][\size]{0.2}
\psfrag{15}[c][c][\size]{0.3}\psfrag{0.2}[c][c][\size]{1}
\psfrag{0.25}[c][c][\size]{2}
\psfrag{777}[c][c][\size]{$\omega$[rpm]}
\includegraphics[height=3cm,width=9cm]{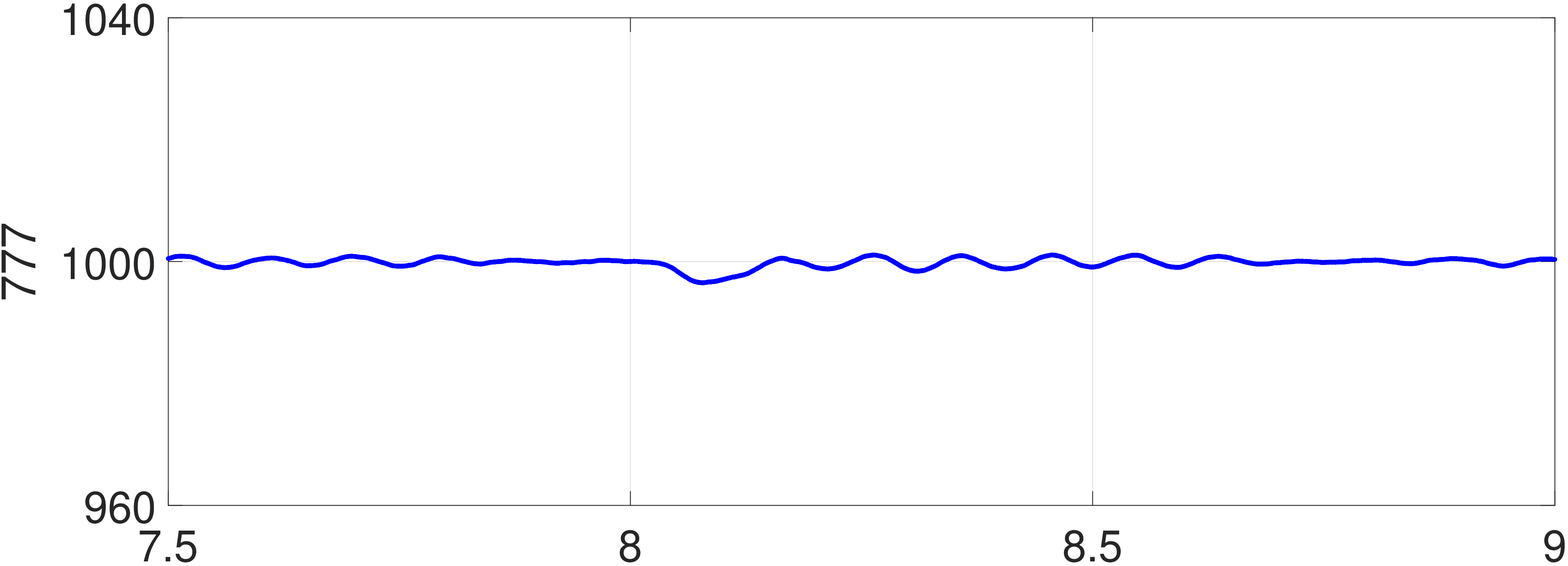}
\psfrag{9}[c][c][\size]{0}
\psfrag{888}[c][c][\size]{$T_e$[N$\cdot$m]}
\psfrag{555}[c][c][\size]{Time[s]}
\includegraphics[height=3cm,width=9cm]{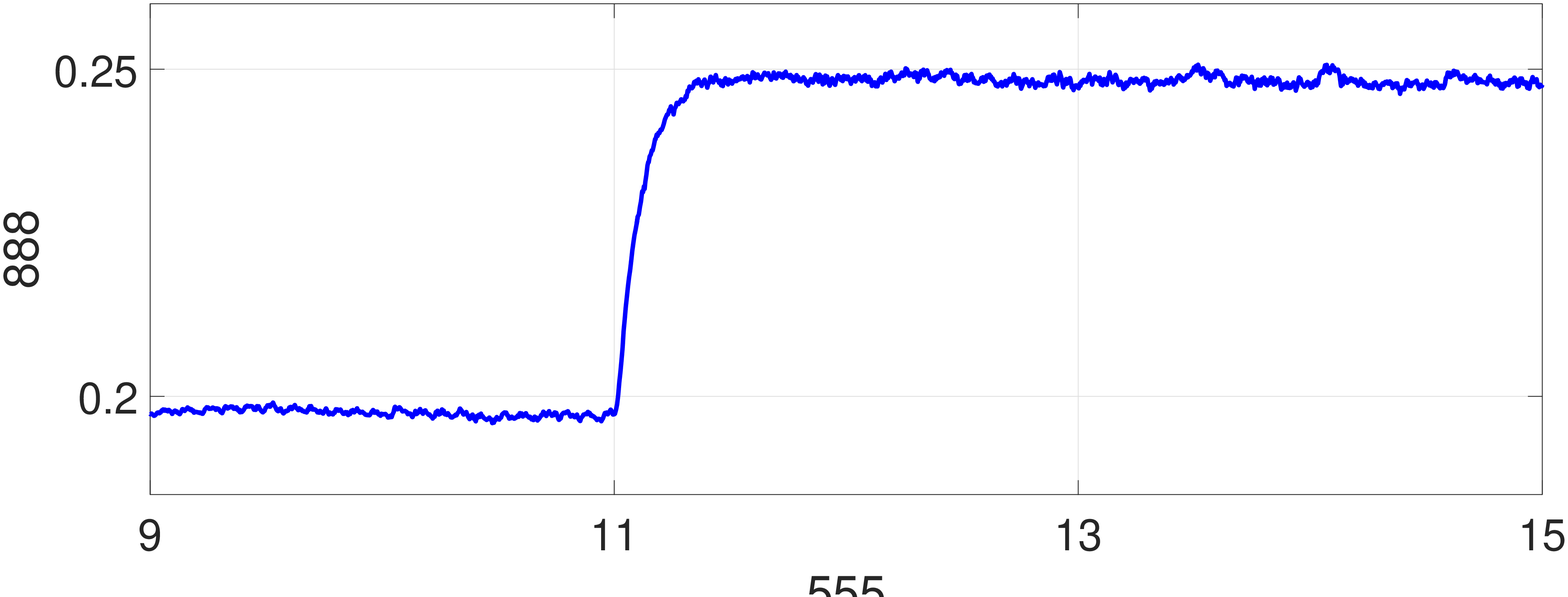}
\begin{center}(b)
\end{center}
\end{center}
\caption{(Experiment). Response curves for dynamic performance evaluation under load disturbance torques: (a) IM MPCC without DC; and (b)
IM MPCC with DC.}
\label{1}
\end{figure}

\begin{table}[htbp]
\caption{Comparison of Performance indicators (Experiment)}
\begin{center}
\begin{tabular}{|c|c|c|c|c|}
\hline
{\textbf{Control Strategy}}&\multicolumn{4}{|c|}{\textbf{Performance indicators}} \\ \cline{2-5}
&\multicolumn{2}{|c|}{\textbf{$\ e_{max}$}}& \multicolumn{2}{|c|}{\textbf{$\ t_{c}$}} \\ \cline{2-5}
&\textbf{speed}& \textbf{torque}& \textbf{speed}& \textbf{torque} \\
\hline
IM MPCC& -8.32$rpm$& -& 0.23$s$& 0.045$s$ \\
IM MPCC+DC& -4.19$rpm$& -& 0.15$s$& 0.022$s$ \\
\hline
\end{tabular}
\label{tab1}
\end{center}
\end{table}

\subsection{THD }

The THD values of three-phase current with the fundamental wave frequency of 50 Hz are compared in Table 4. As compared with MPCC, the THD of IM MPCC strategy decreased by 22.03\% in average, which indicates that IM MPCC strategy has much better steady-state performance.
\begin{table}[htbp]
\caption{Three-phase Current THD Comparison Data (Experiment)}
\begin{center}
\begin{tabular}{|c|c|c|c|}
\hline
\textbf{Control}&\multicolumn{3}{|c|}{\textbf{Three-phase Current THD}} \\
\cline{2-4}
\textbf{Strategy} & \textbf{\textit{$\ i_a$}}& \textbf{\textit{$\ i_b$}}& \textbf{\textit{$\ i_c$}} \\
\hline
MPCC& 10.552$\ \rm{\% }$& 10.519$\ \rm{\% }$& 10.538 $\ \rm{\% }$ \\
IM MPCC& 8.012$\ \rm{\% }$& 8.133$\ \rm{\% }$& 8.087$\ \rm{\% }$ \\
\hline
\end{tabular}
\label{tab1}
\end{center}
\end{table}

\section{CONCLUSION}
In this paper, an improved multi-step FCS-MPCC strategy with speed loop disturbance compensation for PMSM drives has been investigated. In order to improve the steady-state performance and reduce computation burden, an IM MPCC strategy is proposed. To further improve the dynamic response of  the closed-loop system, a disturbance compensation (DC) mechanism based on extended state observer (ESO) has been added to estimate and compensate the total disturbance of the speed loop for PMSM system before the IM MPCC controller. The simulation and experimental results show that the steady-state performance and dynamic response are both improved in the proposed strategy. Therefore, the proposed control strategy in this paper has a good application prospect for high-precision  PMSM drives and easier to implement.

\end{document}